# Nanoindentation creep of supercrystalline nanocomposites

Cong Yan[1], Büsra Bor[2], Alexander Plunkett[2], Berta Domènech[2], Verena Maier-Kiener[3], Diletta Giuntini[1,2,*]

[1] Department of Mechanical Engineering, Eindhoven University of Technology, The Netherlands

[2] Institute of Advanced Ceramics, Hamburg University of Technology, Germany

[3] Department of Materials Science, Montanuniversität Leoben, Austria

**Abstract**

Supercrystalline nanocomposites (SCNCs) are inorganic-organic hybrid materials with a unique periodic nanostructure, and as such they have been gaining growing attention for their intriguing functional properties and parallelisms with hierarchical biomaterials. Their mechanical behavior remains, however, poorly understood, even though its understanding and control are of paramount importance to allow SCNCs' implementation into devices. An important aspect that has not been tackled yet is their time-dependent deformation behavior, which is nevertheless expected to play an important role in materials containing such a distribution of organic phase. Hereby, we report on the creep of ceramic-organic SCNCs with varying degrees of organic crosslinking, as assessed via nanoindentation. Creep strains and their partial recoverability are observed, hinting at the co-presence of viscoelasticity and viscoplasticity, and a clear effect of crosslinking in decreasing the overall material deformability. We rationalize our experimental observations with the analysis of stress exponent and activation volume, resulting in a power-law breakdown behavior and governing deformation mechanisms occurring at the organic sub-nm interfaces scale, in terms of organic ligands rearrangement. The set of results is reinforced by the evaluation of the strain rate sensitivity via strain rate jump tests, and the assessment of the effect of oscillations during continuous stiffness measurement mode.

[*] Corresponding author: d.giuntini@tue.nl

**Introduction**

Nanocomposites are a material category that is particularly promising when it comes to fostering new and targeted combinations of functional and mechanical properties, but their controlled processing remains a challenge [1]. Supercrystalline nanocomposites (SCNCs) show great potential in this regard, since their peculiar nanostructure offers unique opportunities for the fine-tuning of their design at the nano- and micro-scale, and thus of their macroscopic behavior [2,3]. SCNCs consist of inorganic nanoparticles (NPs) that are surface-functionalized with organic ligands and architected into long-range ordered structures. These are reminiscent of atomic crystal lattices, even though at a characteristic scale that is typically two orders of magnitude larger, and are thus called superlattices [3]. Two key strategies to achieve these nanostructures (often of the face- or body-centered cubic type, FCC or BCC) are the organic functionalization of the NP cores, which prevents their uncontrolled aggregation during processing, and a narrow size distribution [4,5]. The combination of nano-sized building blocks and their periodic arrangement in superlattices with ultra-thin inter-NP spacings leads to a wide spectrum of emergent collective properties – electronic, optical, magnetic, plasmonic, thermal, and more. Furthermore, SCNCs present a strong analogy with biomaterials, due to their nanostructured and hierarchical architectures consisting of a tailored mix of inorganic-organic constituents, and one can thus foresee potentially achieving enhanced mechanical properties for applications in e.g. bioimplants, additional to those in battery electrodes, catalysts and optoelectronic devices enabled by their functional properties [3,6–10].

In all these instances, then, understanding and controlling the mechanical behavior of this new material category acquires paramount importance. The mechanical aspects of SCNCs are, however, still largely unexplored. Among the several factors affecting their mechanical properties (NP size [11–13], packing density [13], ligand-NP binding strength, interactions among ligands [11,13], and between ligands and solvent the NP system is stabilized into [14]), the organic ligands themselves play a key role. By inducing their crosslinking, for instance, it has been shown that a remarkable increase in the SCNCs strength, hardness, stiffness, and even fracture toughness, can be achieved [4,5,11–19]. On the other hand, the presence of an organic phase leads to the presence of time-dependent features in their mechanical response, an aspect that has been detected but not analyzed so far [17,20]. This is expected also considering again the analogy with organic-mineral biomaterials, for which creep has been observed and characterized in several instances (e.g. enamel [21,22], bone [23], intervertebral disc [24]). Creep occurs when testing SCNCs [17], but the underlying deformation mechanisms remain to be unveiled.

Uniaxial creep testing is relatively cumbersome to implement when it comes to SCNCs, since their production in large sizes is still quite challenging. Nevertheless, creep testing via instrumented indentation has progressively been perfected since its initiation in the 1970s [25], and nanoindentation has become a reliable, efficient and well-established method to evaluate the creep behavior of nano- and micro-structured materials [26], such as nanocrystalline metals [27–29], ceramics [[30,31], amorphous materials [32], polymers [33] and biomaterials [21–24]. Even though the stress field's state and magnitude induced via indentation-based testing can lead to deviations from uniaxial testing [26,34], many analyses have however shown a very good agreement between the creep behavior measured with the two testing methods [27,32,35].

There is an additional important aspect to consider when it comes to nanoindentation creep. Although indentation creep tests are typically carried out under single-loading mode (i.e. quasi-static load), they can also be conducted in the so-called continuous stiffness measurement (CSM) mode [36]. Under CSM mode, a harmonic load with small amplitude and high frequency is superimposed on the quasi-static load, enabling a continuous estimation of contact stiffness, and thus a more accurate estimation of elastic modulus and hardness, additionally reducing the influence of thermal drift, which represents a major concern in creep studies [37–40]. These super-imposed oscillations can, however, lead to alterations of the measured mechanical behavior [41,42]. Potential sources of error have been identified (underestimation of applied load, displacements or contact stiffness, and loss of contact in the initial test stages) and both preventive strategies and corrections have been proposed [43] – such as not applying this method to materials with large $E/H$ ratios to avoid alterations in the contact stiffness due to local plasticity [44]. Nevertheless, a great variety of deformation phenomena can still occur during CSM-based tests, such as fatigue-associated failure, softening associated with dislocation nucleation in metallic materials, or hardening associated with strain rate sensitivity [45–48]. When it comes to SCNCs, given the peculiarity of their tightly packed arrangement and the extreme confinement of the organic ligands (in the same length-scale of the harmonic oscillations), the effect of the CSM testing mode cannot be directly associated with any of these mechanisms, and thus calls for a separate analysis.

Here, the creep behavior of SCNCs is characterized and analyzed under quasi-static and CSM loading modes. The deformation is assessed for both primary and secondary creep regimes, and for recoverability. The classic concepts of correlation between stress exponent (also tested via strain rate jump tests), activation volume and creep mechanisms are critically applied to this new material category.

**Materials and methods**

**Samples preparation**

The supercrystalline nanocomposites consist of iron oxide (magnetite, $Fe_3O_4$) NPs, surface-functionalized with oleic acid (OA) and initially suspended in toluene (Fraunhofer CAN GmbH, Hamburg, Germany). Starting from these building blocks, the formation of the supercrystalline structure is induced by self-assembly via solvent destabilization [17,18]. The NP suspension (40 mg/mL) is poured into a die-punch assembly (14 mm diameter), and placed into a desiccator, in which the atmosphere is then enriched with ethanol, acting as destabilization agent upon slow diffusion into the NP colloidal suspension. The self-assembly process lasts ~15 days. The sedimented SCNCs ae then recovered by removal of the supernatant with a pipette. The samples are dried for 24 h under ambient conditions followed by 2 h under vacuum. By means of a second punch they are pressed uniaxially (in the rigid die) at 150 ºC and with 50 MPa, to obtain bulk cylindrical pellets, ~4 mm thick. The pellets are then cut into three portions. One is left as-pressed (AP), while the other two are subjected to a heat treatment (HT) at 250 and 325 ºC under nitrogen ($N_2$) atmosphere (heating and cooling ramps 1 ºC/min, holding time 18 min). The heat treatment induces the crosslinking of the organic ligands, and in turn a boost of the SCNCs' mechanical properties [4,17–19,49]. The different types of samples (AP, HT250, HT325) are then fixed on SEM stubs with silver glue and embedded into cold-curing acrylic resin (Scandiquick, Scan-DIA, Hagen, Germany) to test the cross-section of the pellets. The samples' surface is polished with SiC paper and then diamond suspensions, down to a roughness of 50 nm.

**Composition and nanostructure characterization**

To quantify the organic content in the 3 different materials (AP, HT250 and HT325), thermogravimetric analysis (TGA) is conducted in a Mettler Toledo TGA/DSC1 STARe System (Mettler Toledo, UK), from 25 to 900 °C with a heating rate of 5 °C/min, under nitrogen ($N_2$) atmosphere. The indents are observed by scanning electron microscopy (SEM, Zeiss 55-VP, Zeiss, Germany) at 2 kV, to verify that the indents are in supercrystalline areas (avoiding defects) and that severe damage has not been induced around the indents. Topographic measurements are conducted via atomic force microscopy (AFM, NanoScope IV, Dimension 3100 of Digital Instruments, USA), with a 0.5 Hz scanning speed.

**Nanoindentation**

The nanoindentation tests are performed in an G200 system (KLA, formerly Agilent, USA) at room temperature, with a diamond Berkovich tip (Synton-MDP, Switzerland), chosen for its geometrical self-similarity [29] and wide use in creep tests. The single-loading mode (quasi-static) tests consist of five phases: I) loading; II) creep; III) unloading; IV) backcreep (a low-load, 2% of the maximum load, holding step aimed at investigating creep recoverability); V) thermal drift measurement (see the inset of Fig. 2 (a)). For I and III, a rate 0.2 mN/s is selected, to guarantee a fast loading/unloading, as well as to avoid overshooting the prescribed maximum load. During II and IV, the holding time is 1000 s. The maximum loads are 2.5, 4, 5.5 and 10 mN, corresponding to 200-500 nm depths, to ensure that the indenter penetrates the sample for a number of layers of nanoparticles representative of the bulk material, as well as to avoid the occurrence of cracking and chipping damage, based on preliminary tests and previous work [17]. The choice of conducting tests in load control mode instead of displacement control was due to the better assessment of thermal drift that load-control allows, but tests in displacement-control mode were also conducted to check on the general applicability of the results.

The CSM loading mode is also employed in this study, for 2 purposes: i) measuring the SCNCs' elastic modulus ($E$) and hardness ($H$) (constant strain rate of 0.05 $s^{-1}$ and maximum depth of 300 nm [17], still Berkovich tip); ii) investigating the influence of oscillations on the creep deformation, with a harmonic amplitude of 2 nm and a frequency of 45 Hz. The tests under CSM mode consist of the same 5 steps.

At least 12 indents were performed for each load and measurement mode, at a distance of 30 µm from each other. To eliminate the effect of thermal drift during long-term indentation experiments, a multi-approach analysis is conducted, as explained in Appendix A. The tip calibration is performed on silica.

**Results and Discussions**

**1. Nanostructure, composition and mechanical properties of supercrystalline nanocomposites**

The starting constituents and the final nanostructure of the SCNCs are characterized by transmission and scanning electron microscopy (TEM, SEM) and small angle X-ray scattering (SAXS), see SI section S1 and Fig. S1. The NP radius is 7.4 ± 0.8 nm according to SAXS (Fig. S1(c)). SAXS also reveals the face-centered cubic (FCC) superlattice of the self-assembled SCNCs for all three types of samples (Fig. S1(d)), with superlattice constants of 22.8 ± 0.04 nm for AP, 22.9 ± 0.03 nm for HT250 and 21.9 ± 0.12 nm for HT325, corresponding to interparticle distances (*ID*) of 1.2 ± 0.03 nm, 1.3 ± 0.02 nm and 0.6 ± 0.09 nm, respectively, as reported in Table 1. Considering the length of the totally extended OA molecule (~2 nm), these values suggest that the organic ligands are interdigitated and/or bent at the interfaces between NPs. The organic content is measured by TGA, as summarized in the same Table 1 (see also SI Section S2 and Fig. S2). Superlattice shrinkage, typically reported for heat-treated SCNCs [49], is only observed in the HT325 case, while the HT250 samples have only a minor reduction of the organic content with respect to the AP counterparts. This is because the reactions induced by the heat treatment and that lead to strengthening and ID reduction proceed at different rates, and thus at 250 ºC one detects strengthening without significant shrinkage, while at 325 ºC both are detected [19]. Based on the TGA-measured organic content in AP samples, the grafting density of the OA on the NP surface is also obtained, as ~2.4 molecules/nm$^2$ (see SI Section S3). This is then compared with the maximum theoretical grafting density, i.e. the amount of organic ligands required to form a full monolayer on the NP surfaces. Assuming that the ligands are oriented perpendicular to the NP surface and that the entire NP surface is available for the binding of ligands, the theoretical grafting density is then determined as 4.8 molecules/nm$^2$ (considering that the area of the OA anchoring group to the NP is 0.21 nm$^2$) [50]. It should be noted that this theoretical value provides an upper bound of grafting density, since the NP surfaces are likely to have binding sites that do not allow their complete coverage with anchored ligands, and that at the same time TGA can underestimate the overall organic content, due to the fact that some residual carbon can remain in the measured samples, on the NP surface [51]. Nevertheless, the theoretical value is much larger than the experimental one, making it reasonable to assume that the superlattice interstitial sites are empty.

**Table 1.** Superlattice parameters (superlattice constant and interparticle distance), organic content and mechanical properties (elastic modulus and hardness) of the tested supercrystalline nanocomposites (SCNCs).

|  | Superlattice constant, $a$ (nm) | Interparticle distance, $ID$ (nm) | Organic content (wt%) | Elastic modulus, $E$ (GPa) | Hardness, $H$ (GPa) |
|---|---|---|---|---|---|
| AP | 22.8 ± 0.04 | 1.2 ± 0.03 | 8.0 | 37.9 ± 2.6 | 1.97 ± 0.24 |
| HT250 | 22.9 ± 0.03 | 1.3 ± 0.02 | 7.3 | 43.1 ± 4.2 | 2.40 ± 0.38 |
| HT325 | 21.9 ± 0.12 | 0.6 ± 0.09 | 5.5 | 59.8 ± 6.2 | 4.19 ± 0.52 |

The SCNCs' elastic modulus and hardness are measured via nanoindentation in CSM mode (see SI Section S4 and Fig. S3) and the corresponding values are also summarized in Table 1. With increasing heat treatment temperature, crosslinking is induced, and covalent bonds are formed in-between adjacent OA molecules, leading to a high-strength network holding the material together [4,19]. Therefore, the elastic modulus and hardness are gradually enhanced, and this increase is particularly remarkable for the HT325 sample (58% increase of elastic modulus and 113% increase of hardness with respect to the AP case). For the HT250 sample, the enhancement is less pronounced, as a result of slight decrease of organic content and less pronounced crosslinking. These $E$ and $H$ values are remarkably high for ceramic-organic nanocomposites, and especially for supercrystalline materials, which usually feature the elastic moduli in the 1 – 20 GPa range and hardness around 40 – 450 MPa [12].

The typical SCNCs nanostructure is shown in Fig. 1(a), displaying a fracture surface from the AP material, which is representative of all samples (the differences among them are not detectable at this resolution). The long-range order arrangement of the NPs is visible. The superlattice orientation is not uniform throughout the cm-sized samples, thus rendering the SCNCs poly-supercrystalline, and the probed mechanical properties can in turn be considered isotropic. The morphologies of indents obtained with a 5.5 mN load are shown below, in Fig. 1(b) – (d). All indents are analyzed at the SEM, confirming that the wide majority have been performed in supercrystalline domains, and not in localized defect-affected or amorphous regions. The varying superlattice orientations can also be observed from these same micrographs. Indentation-induced damage, such as cracking and chipping, is not detected in the indents performed with 5.5 mN, and it is then reasonable to conclude that no damage is induced by indenting at lower loads (i.e. 4 mN and 2.5 mN). The indents obtained at 10 mN are also examined, see SI Fig. S4, whereby slight damage can only be seen in a few indents in AP and HT250 samples, but none in the HT325 sample. The data relative to the 10 mN load is thus also considered in the following analysis.

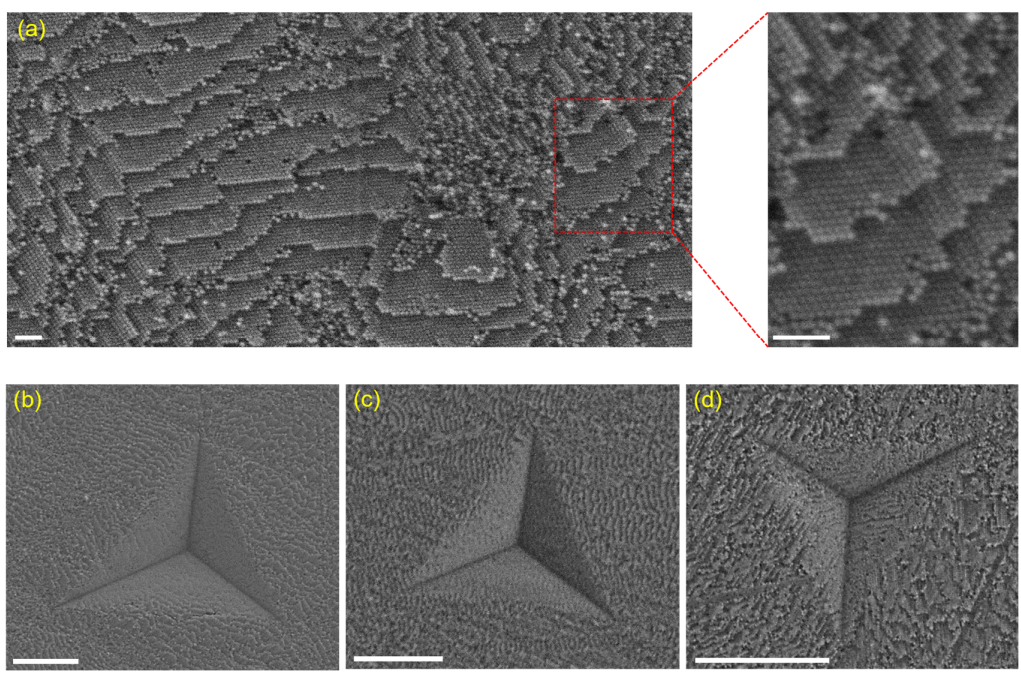

**Fig. 1.** Nanostructure and indents in SCNCs. (a) Fracture surface in AP sample, with clear view of the periodic NP arrangement in the magnified image; (b) – (d) indents in different samples obtained with 5.5 mN load: b) AP; c) HT250; d) HT325. All images are SEM micrographs. Scale bars: 100 nm in (a), 1 μm in (b) – (d).

**2. Creep occurrence**

Load-displacement nanoindentation curves obtained in 2.5, 4, 5.5 and 10 mN tests are shown in Fig. 2(a) – (d), respectively. For all applied loads, the maximum displacements gradually decrease from AP to HT325 SCNCs, as expected thanks to the crosslinking-associated enhancement of the mechanical properties. Creep occurs during the holding at maximum load, while recovery occurs during the backcreep stage, as marked in Fig. 2(c). To compare the evolution of creep and backcreep displacements of different samples, these are displayed for the 5.5 mN case in Fig. 2(e) and (f), respectively, and in Fig. S5 for the three other loads. In Fig. 2(e), one can observe that creep displacements first increase quickly and then slow down. The higher the heat-treatment temperature, the smaller the overall creep displacements, indicating that creep resistance is enhanced by the organic ligands' crosslinking. The backcreep displacements in Fig. 2(f) show a similar trend, by initially dropping sharply and then slowing down. The tests under different loads (Fig. S5) share the same features.

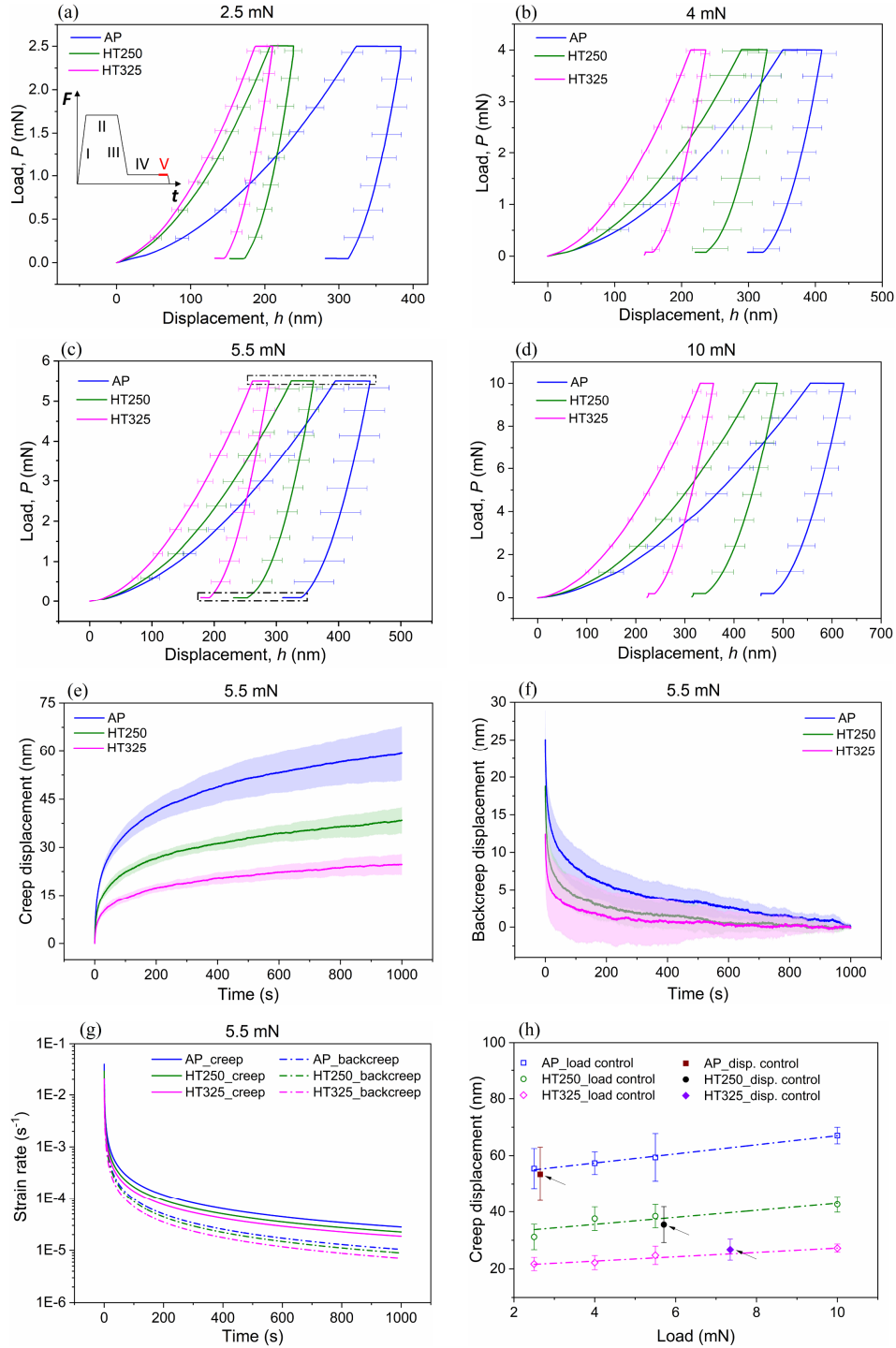

**Fig. 2.** Analysis of nanoindentation creep deformation of SCNCs. (a) – (d) Load-displacement curves under different loads: (a) 2.5 mN; (b) 4 mN; (c) 5.5 mN; (d) 10 mN. The inset in (a) displays the loading scheme (I – loading tamp; II – creep; III – unloading ramp; IV – backcreep; V – thermal drift measurement); (e) – (f) Evolution of creep/backcreep displacements in 5.5 mN tests (marked by dashed boxes in (c)): (e) creep; (f) backcreep. The shading bands indicate the standard deviation. (g) Strain rates in 5.5 mN tests during creep and backcreep; (h) Comparison of creep displacements as obtained in load-control (hollow symbols) and displacement-control mode (solid symbols, indicated by arrows) creep tests.

Strain rates are also analyzed, to evaluate the difference of deformation rate between creep and backcreep. The empirical power-law expression $h(t) = a + b * t^c$ is adopted to fit the results [38], and the strain rate is then calculated as:

$$\dot{\varepsilon} = \frac{\dot{h}}{h} = \frac{dh/dt}{h} \tag{1}$$

where $h$ is the total displacement [32]. The evolution of strain rates during creep and backcreep is shown in Fig. 2(g). The strain rates decay quickly, both for creep and backcreep, and tend to a quasi-constant value, with the backcreep strain rates consistently smaller than those of creep, by a factor of ~3. As typical in nanoindentation tests, due to the multiaxial stress, a fully steady state (constant strain rate) is not reached [26], so we will refer to this later creep stage as quasi-secondary (or quasi-steady state).

The varying maximum displacements in the different materials (Fig. 2(a) – (d)) indicate that for each case different volumes are involved by the indentation tests when the same load is applied. To verify whether the volume of deformed material impacts creep deformation, displacement-control (300 nm) creep tests were also conducted, keeping all other parameters unchanged. It should be noted that previous works have revealed the occurrence of compaction in SCNCs [17,49], implying that even in displacement-control tests one cannot attain exactly equivalent volumes of material affected by the indentation load, but it is still reasonable to assume that such volumes are sufficiently close to each other when the indentation depth is kept constant. The comparison of creep displacements between load-control and displacement-control creep tests is summarized in Fig. 2(h). The creep displacements under load-control tests exhibit a linear relationship with the load, with which the data from displacement-control tests is aligned, indicating that no significant difference can be detected between load-control and displacement-control creep tests, and that the influence of different indented volumes can thus be neglected.

## 3. Recoverability of creep deformation

Due to the presence of an organic phase in the SCNCs, and the typically associated viscous material behavior, creep recovery, at least partial, is analyzed. The same phenomenon has indeed been consistently observed and noticed in biomaterials (enamel [21], intervertebral disc [24]) and polymers [52,53]. For instance, He et al. reported that almost 40% of creep deformation in enamel is recovered during backcreep, indicating that both viscoelastic and viscoplastic deformation occur [21], while certain polymers can achieve complete recovery [52]. The displacements during creep and backcreep under the four different loads are compared in Fig. 3(a), where one can see that backcreep displacements are consistently smaller than creep ones, with the ratios of backcreep to creep displacements ranging from 40% to 50%, similarly to the values reported in [21], thus indicating that creep deformation does not fully recover during backcreep. This only partial recovery is associated with a relatively low organic content, and especially the high confinement of the organic ligands in ultra-thin inter-NP spacings, in which the OA molecules are anchored to the NP surfaces, likely interdigitated and sometimes also crosslinked.

From Fig. 2(g) one can notice, however, that the strain rate during backcreep does not reach zero, and it might thus keep proceeding after nanoindentation. AFM measurements are then performed several weeks after the nanoindentation tests, on selected indents, to verify whether the recovery continues after the indenter's withdrawal.

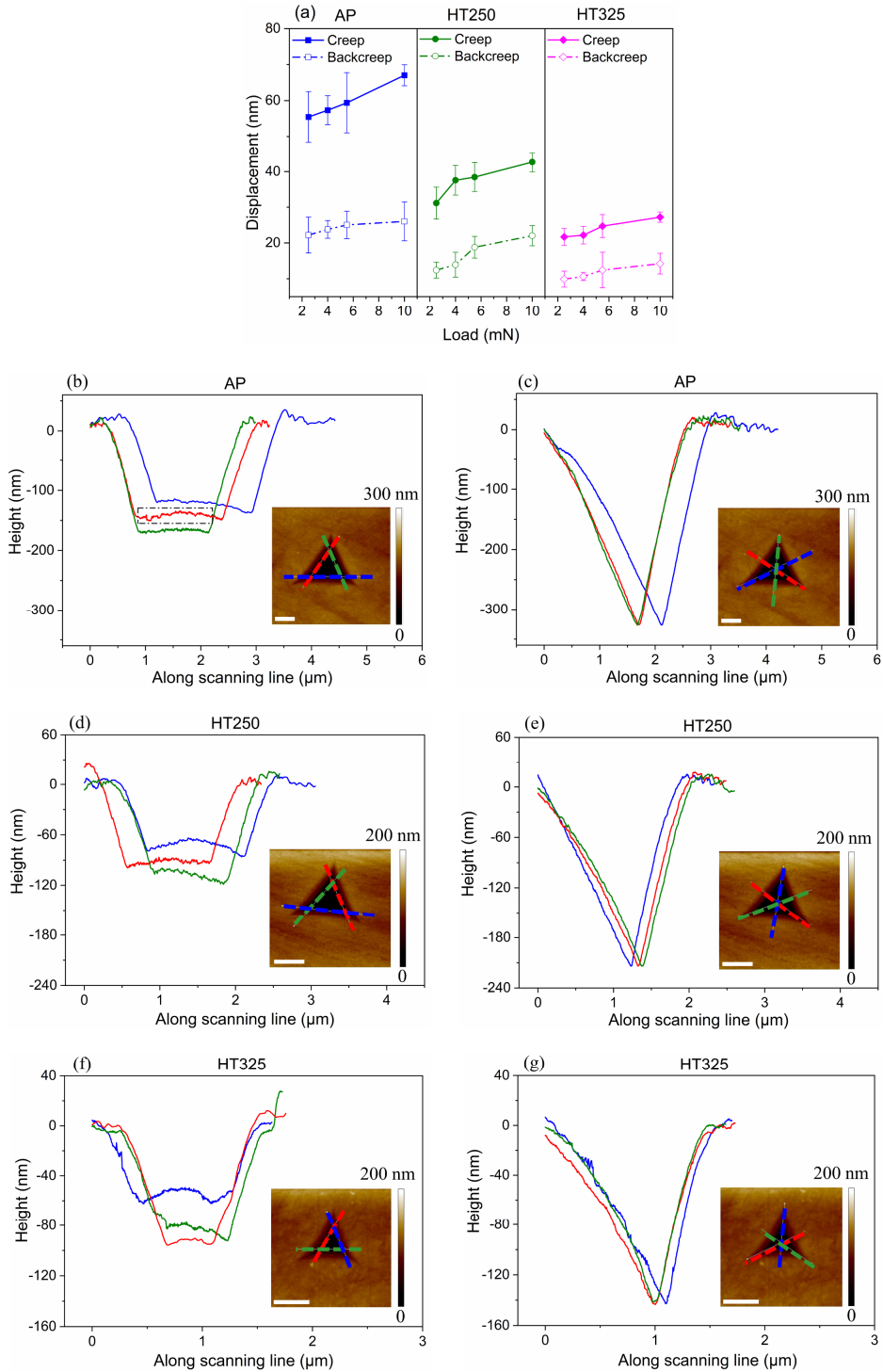

**Fig. 3.** Creep recovery analysis via nanoindentation and AFM. (a) Creep and backcreep displacements under different loads, as recorded by the nanoindenter. (b) – (g) Depth profiles along corresponding scanning lines in AFM topography maps, verifying the occurrence of recovery after nanoindentation: (b) – (c) AP sample; (d) – (e) HT250 sample; (f) – (g) HT325 sample. The bottom sections of the profiles in (b), (d), (f) display a somewhat round morphology (marked by a dashed box in (b)), hinting at the occurrence of recovery. The profiles in (c), (e), (g) show the depth variation along different scanning lines in AFM topography maps, adopted to determine the residual depth after nanoindentation, which is then compared with the depth recorded by nanoindentation to quantify the overall recovery. Scale bar is 1 μm in the AFM topography maps.

Fig. 3(b) – (g) show the resulting topography maps and depth profiles. The "bumpy" surfaces (Fig. 3(b), (d), (f)) along the side faces of the indents suggest that recovery continues after the tip is retracted (see e.g. the area marked by the dashed box in Fig 3(b)). The residual depths measured via AFM (Fig. 3(c), (e), (g)) are then compared with those resulting from nanoindentation. For AFM, the intersection of the three scanning lines shown in Fig. 3(c), (e) and (g) is taken as residual depth marker. The difference between residual depth (AFM) and final nanoindentation depth is the displacement recovered after nanoindentation, as summarized in Table 2 for selected representative indents. The recovered displacements are still smaller than creep displacements, indicating that recovery is not complete even several weeks after nanoindentation. It thus appears that viscoelasticity and viscoplasticity both occur during creep, although visco-plasticity only accounts for 20–30% of the overall deformation. The Kohlrausch-Williams-Watts (KWW) model, usually adopted to describe viscoelastic stress evolution during constant load holding [54], also fails to describe the stress evolution in the case of SCNCs. Instead, an expression with double exponential terms describes well this stress evolution ($R^2 > 0.99$), confirming that creep deformation is not solely viscoelastic, and that viscoplastic deformation also occurs (see SI Section S7 and Fig. S6).

**Table 2.** Creep deformation recovery based on nanoindentation and AFM.

|  | Recovery after nanoindentation (nm) | Recovery during backcreep (nm) | Total recovery (nm) | Creep displacements (nm) | Total recovery (%) |
|---|---|---|---|---|---|
| AP | 28 | 20 | 48 | 60 | 80% |
| HT250 | 16 | 12 | 28 | 40 | 70% |
| HT325 | 1 | 16 | 17 | 21 | 81% |

Note: Total recovery = recovery during backcreep + recovery after nanoindentation. Nanoindentation has a resolution of 0.01 nm while AFM has a 0.1nm resolution. Here we rounded both up to integer values.

## 4. Primary creep

By integrating the strain rates obtained from Eq. (1), the strain evolution during holding can be obtained. The outcomes are shown in Fig. 4(a) – (d) (note that the initial strain is set back to zero to focus on creep strains only). The same trend can be found in all plots: the largest strains occur in AP SCNCs, the smallest ones in HT325 samples, with intermediate values in HT250 samples, verifying that crosslinking enhances resistance against creep. An attempt is made to identify the transition point between primary and quasi-steady-state (secondary) creep on the strain-time curves. Based on the definition of steady-state creep, strain should have a linear relationship with respect to time. A linear fitting is thus performed from the end of the strain-time curves, and the transition is set on the point where the linear fitting values deviate from the experimental ones by 2% [55], as indicated by the star symbol on the curves. We can then separately analyze primary and quasi-secondary creep, respectively.

The deformation during primary creep usually proceeds quickly before reaching steady-state creep [33]. If the strain during primary creep exceeds $2\times10^{-3}$, the evolution of strain can be described by a power-law model, i.e. $\varepsilon = Bt^a$, where $B$ and $a$ are constants depending on the material, with the time exponent $a$ indicating the decay rate of the creep strain [56]. This model is applied to the primary creep data, and it is found that the experimental results match the model very well, which also implies that the method to determine the transition point is reasonable. The deformation during primary creep is much larger than that associated with the following creep stage, due to its high deformation rate [33]. More than 80% of the deformation is observed during primary creep, see SI Section S8 and Table S1, indicating that primary creep dominates the total creep deformation.

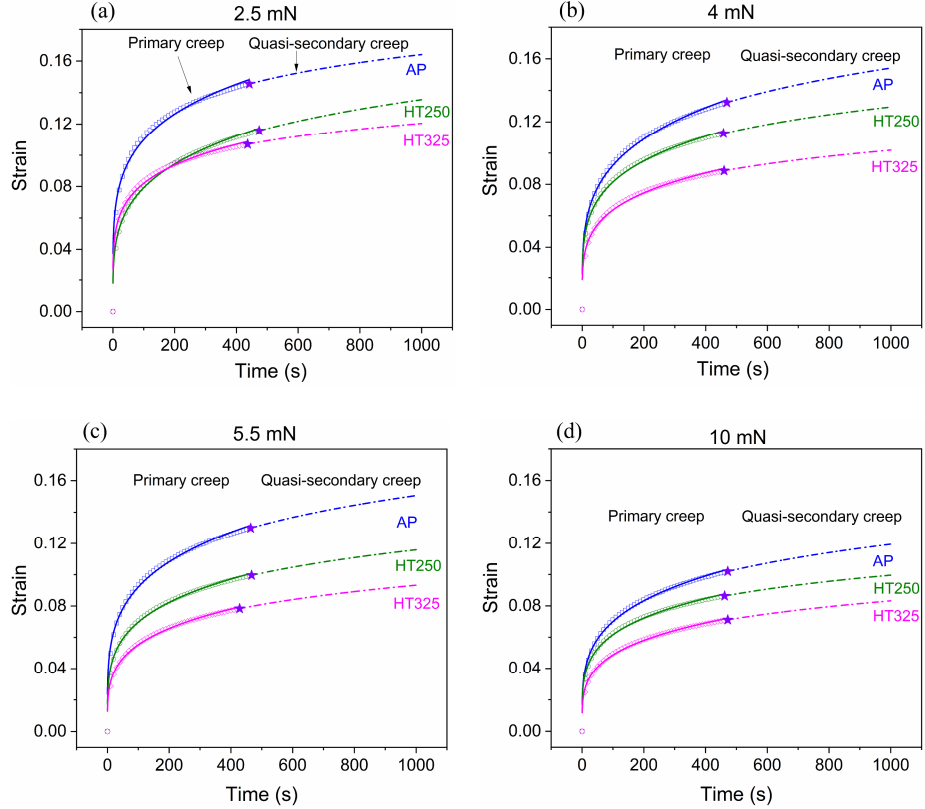

**Fig. 4.** Identification of primary and quasi-secondary (steady-state) creep on the strain-time curves: (a) 2.5 mN; (b) 4 mN; (c) 5.5 mN; (d) 10 mN. The first segment of the curves, plotted as scatter squares, represents primary creep, while the second segment of curves, plotted in dashed lines, marks the quasi-secondary creep. The shift between the two regimes is marked by transition points, indicated by star symbols. A power-law model fits the strain during primary creep (solid lines) with a high correlation coefficient ($R^2 > 0.98$).

## 5. Quasi-secondary creep and underlying mechanisms

Even though accounting for a smaller part of the overall creep deformation, the secondary (steady-state) stage can shed light on the mechanisms of creep deformation. The strain rate and stress during secondary creep are usually correlated via an empirical power-law equation:

$$\dot{\varepsilon} = A\sigma^n \exp\left(-\frac{Q}{RT}\right) \qquad (2)$$

where $\dot{\varepsilon}$ is strain rate, $A$ is a constant associated with microstructural features, $\sigma$ is the stress, $n$ is the stress exponent, $Q$ is the activation energy of creep deformation, $R$ and $T$ are the ideal gas constant and absolute temperature, respectively [28,57]. It is widely accepted that the stress exponent ($n$) is an indicator of creep mechanisms, i.e. $n = 1$ for diffusion creep, $n = 2$ for grain boundary sliding, $n = 3 - 7$ for creep related to dislocation activity, $n > 7$ for power-law breakdown [57–59].

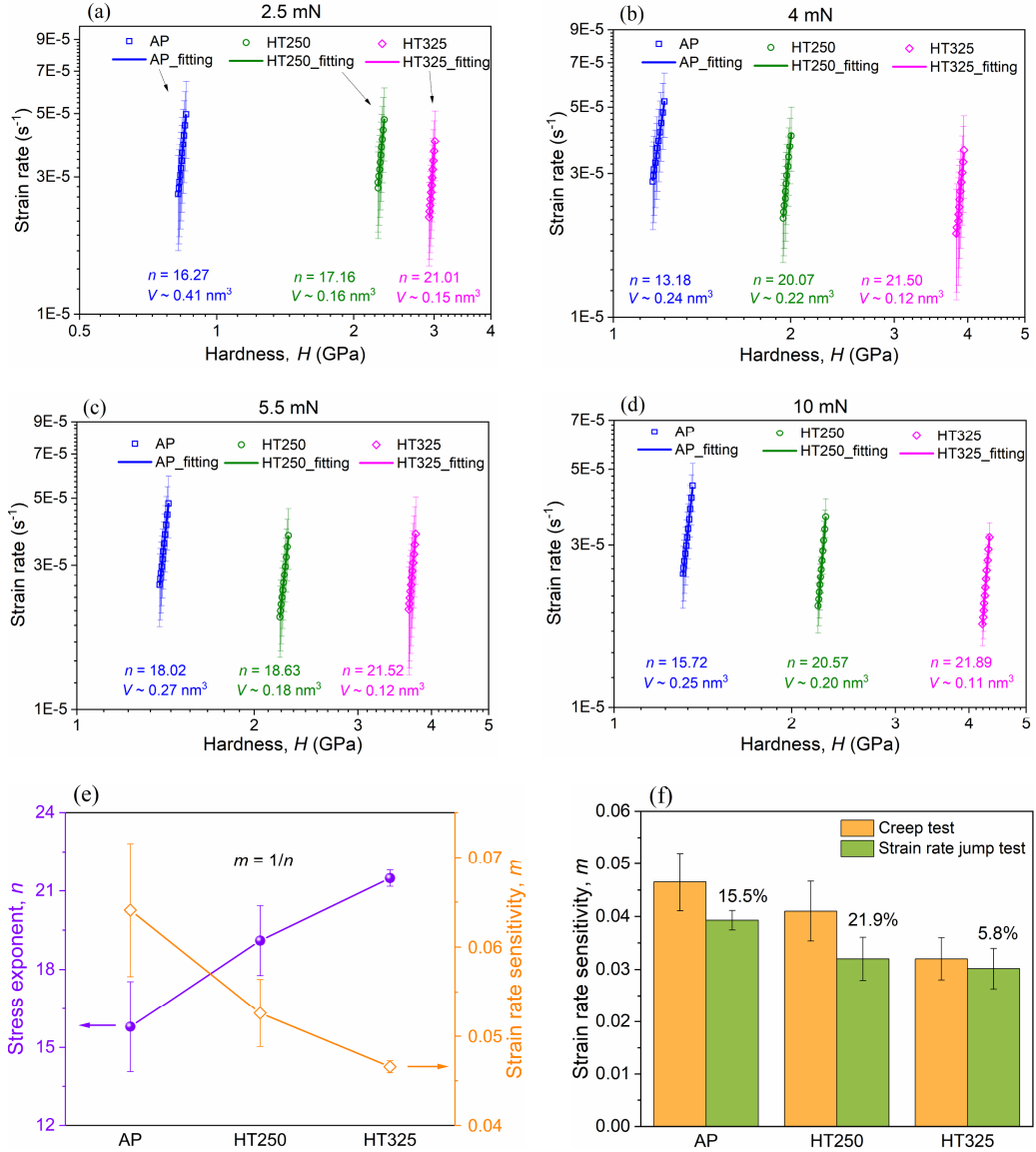

**Fig. 5.** Relations of logarithmic strain rate vs. logarithmic hardness during creep holding, the slope of which corresponds to the stress exponent ($n$): (a) 2.5 mN; (b) 4 mN; (c) 5.5 mN; (d) 10 mN; The activation volume ($V$) is also marked in the plots, corresponding to the slope of logarithmic strain rate vs. linear hardness as suggested by Eq. (5). The relevant plots can be found in Fig. S8. (e) Comparison of the averaged $n$ over the different loads for the three types of SCNCs, with the corresponding strain rate sensitivity ($m = 1/n$). (f) Comparison of strain rate sensitivity ($m$) obtained via creep and strain rate jump tests. The % indications above the columns indicate the difference of $m$ as measured in these two tests, both carried out under CSM mode.

Several previous studies have revealed that the stress exponent ($n$) typically coincides in the cases of nanoindentation and uniaxial tension creep, suggesting that similar creep mechanisms occur, although a difference can sometimes still exist due to the high stress concentrations associated with nanoindentation [27]. As anticipated above, it has also been reported that in nanoindentation creep a fully steady state is generally not reachable [26], but since the strain rates in this study ($\sim 10^{-5}\,\text{s}^{-1}$) are comparable to or even smaller than several other reported strain rates where Eq. (2) has been applied to derive the stress exponent ($n$) [28,60], it is reasonable to analyze such quasi-steady-state creep mechanisms within this framework.

**Stress exponent and strain rate sensitivity**

The stress exponent ($n$) can be obtained as $n = \frac{\partial ln\dot{\varepsilon}}{\partial ln\sigma} = \frac{\partial ln\dot{\varepsilon}}{\partial lnH}$ from Eq. (2), where $\sigma$ is usually made equivalent to hardness ($H$) data in nanoindentation creep. Hardness is calculated as

$$H = \frac{P}{A_C} \tag{3}$$

where $P$ is the load and $A_C$ is the projected contact area [32]. The function of contact area $A_C$ is shown in the Appendix, Eq. (A4). The stress exponent ($n$) is usually computed by linear fitting of strain rate and hardness in logarithmic plots, as shown in Fig. 5(a) – (d). By averaging the stress exponents obtained in the tests under the four different loads, $n$ is determined as 15.80 for AP SCNCs, 19.10 for HT250 and 21.48 for HT325, all of which are larger than 7, associated with the occurrence of power-law breakdown. Power-law breakdown indicates that the stress exponent is no longer constant with changes in applied stress, increasing with rising stress or strain rate [61]. However, the linearity between $ln\dot{\varepsilon}$ and $lnH$ still appears to be valid in Fig. 5(a) – (d), perhaps due to the limited data range, which does not extend enough to reflect the loss of linearity. Meanwhile, $n$ exhibits an increasing trend with heat-treatment temperature, as shown in Fig. 5(e).

Power-law breakdown is related to the high stresses induced by the nanoindenter, and signals the presence of very high creep strain rates. A threshold value above which power-law breakdown occurs is found in $10^{-3}G$, with $G$ shear modulus of the material [62]. The SCNCs shear modulus ($G$) can be obtained from the Young's modulus ($E$) as $G = E/2(1 + \nu)$, assuming the SCNCs as homogeneous and isotropic. The Poisson's ratio $\nu$ is taken as 0.34, as previously derived via FEM for HT325 sample and here extended to all cases [63]. The representative stress under the indent (assumed as representative average over the hydrostatic and plastic sub-indent zones) is estimated from Tabor's law as $\sigma_r = H/C$, where $C$ is a constraint factor, typically $1 - 3$ depending on which regime (elastic/elasto-plastic/plastic) the deformation is in [64]. Here, $C = 3$ is adopted, and slight changes of $C$ do not significantly impact the estimation. The results are summarized in Table 3. Compared with the threshold values calculated based on $G$, all the estimated representative stresses are larger by at least one order of magnitude, confirming that the occurrence of power-law breakdown is expected.

**Table 3.** Threshold stress values for power-law breakdown onset and representative stress beneath the indents.

|  | Shear modulus, $G$ (GPa) | Threshold value (GPa) | Representative stress (GPa) | | | |
|---|---|---|---|---|---|---|
|  |  |  | 2.5 mN | 4 mN | 5.5 mN | 10 mN |
| AP | 14.1 | 1.41×10$^{-2}$ | 0.27 | 0.39 | 0.46 | 0.44 |
| HT250 | 16.1 | 1.61×10$^{-2}$ | 0.75 | 0.65 | 0.74 | 0.74 |
| HT325 | 22.3 | 2.23×10$^{-2}$ | 0.98 | 1.37 | 1.24 | 1.40 |

Large stress exponents ($n > 7$) have also been reported in creep tests of UFG (ultrafine grained)-Al [37], UFG-Au [38], nanocrystalline Ni [26] and high entropy alloys [29]. All these tests were also performed with a sharp indenter tip. The creep study on the high entropy alloy (CoCrFeMnNi) revealed that a low stress exponent ($n = 1.3$) was obtained from nanoindentation with a spherical indenter, while it sharply rose to 28.5 with a Berkovich indenter [29], confirming that large stress exponent values are usually obtained in nanoindentation creep tests with a sharp tip, such as the Berkovich tip, which can induce high stress concentration beneath the indents.

In general, the stress exponent increases with increasing heat treatment temperature, see Fig. 5(e). In AP SCNCs, the NPs are held together by weak interactions, mainly van der Waals forces [4], making it relatively easy for the NPs to move with respect to each other, also thanks to the OA interfaces acting as a soft and somewhat flexible interlayer. When crosslinking is induced and covalent bonds are formed in heat-treated samples (HT250 and HT325), instead, a higher stress is required to enable the NP movement, accounting for the increasing $n$ values.

Strain rate sensitivity ($m$), the reciprocal of the stress exponent ($m = 1/n$), is an indication of the material's ability to flow viscously. The larger $m$, the more homogeneous the material's flow is [21]. The strain rate sensitivity, shown in Fig. 5(e), decreases with heat treatment temperature, indicating that the flow becomes progressively more inhomogeneous, and that localized shear flow can be present. The declining trend of $m$ is another reflection of the crosslinking effect. Very low $m$ values are associated with a quasi-perfectly-plastic behavior, as indicated in previous TEM studies on SCNCs [49].

The strain rate sensitivity is also re-examined by strain rate jump tests under CSM mode [38,65,66], to verify whether it is comparable to that obtained via creep tests, see Fig. 5(f). Strain rate jump tests rely on lower transition times to reach prescribed strain rate values, ensuring high accuracy in the assessment of $m$. The tests are conducted as follows: the strain rate drops from $2.5 \times 10^{-2}$ s$^{-1}$ to $2.5 \times 10^{-3}$ s$^{-1}$ when the displacement reaches 300 nm, and subsequently rises to $2.5 \times 10^{-2}$ s$^{-1}$ again when the displacement exceeds 400 nm. Strain rate sensitivity is then calculated from [66]

$$m = \frac{\partial ln\sigma}{\partial ln\dot{\varepsilon}} = \frac{\partial lnH}{\partial ln\dot{\varepsilon}} \qquad (4)$$

when the strain rate jumps from the low value to the high one, see Fig. S7(a) & (b). The $m$ obtained from strain rate jump tests is in good agreement, even though slightly smaller, with that from creep tests, likely due to the difference of strain rates (2–3 orders of magnitude). A fairly good agreement between the two methods is also found in the relationship between hardness and strain rates, see SI Section S9 and Fig. S7(c). This comparison confirms that strain rate jump tests are also a viable option to measure the strain rate sensitivity of SCNCs, with the advantage of a procedure that is less time-consuming and less impacted by thermal drift. However, strain rate jump tests are affected by a limitation when even lower strain rates are needed, due to the lower bound of loading rate characteristic of the nanoindenter [39]: a more exact $m$ would be determined under a lower strain rate, because of the requirement of steady-state when applying Eq. (4).

**Activation volume**

The activation volume ($V$) is usually seen as an indicator of the number of atoms involved in the creep deformation process [28], and its applicability is here assessed for the case of functionalized NPs in SCNCs (the crystalline lattice in each NP undergoes negligible strains [67]). This parameter can indeed provide further insight into the creep mechanisms, and it is expressed as

$$V = 3\sqrt{3}kT \cdot \frac{\partial ln\dot{\varepsilon}}{\partial H} \qquad (5)$$

where $\dot{\varepsilon}$ and $H$ are strain rate and hardness, respectively, $k$ is Boltzmann's constant and $T$ is the absolute temperature [29]. By performing a linear fitting of logarithmic $\dot{\varepsilon}$ and linear $H$, activation volumes are determined, as displayed in Fig. S8(a) – (d), with the corresponding values also marked in Fig. 5(a) – (d) for each load. The activation volume reduces with heat-treatment temperature in all tests. The average $V$ values result in 0.25 nm$^3$ for AP SCNCs, 0.19 nm$^3$ for HT250, and 0.13 nm$^3$ for HT325. This trend also suggests that crosslinking hampers creep deformation.

Activation volume, at least in metals and alloys, is correlated with thermally-activated deformation mechanisms [68], such as plastic deformation in nanocrystalline materials [68,69], the nucleation of shear bands in metallic glasses [70], heterogeneous dislocation nucleation to account for incipient plasticity[71]. $V$ is thus usually related to the movement of dislocations in crystalline materials. Given the materials' Burgers vector $b$, $V$ can range from about 1000 $b^3$ (coarse grains) to several $b^3$ (nanocrystalline materials), and it is rationalized as the distance between obstacles during the movement of dislocations [37,66,68]. The activation volume of SCNCs (0.13 nm$^3$ ~ 0.25 nm$^3$) has values comparable to those of nanocrystalline materials (0.1 nm$^3$ for nanocrystalline Ni) [37,68]. But due to the extremely large Burgers vectors at the superlattice scale (here 2 orders of magnitude larger than in atomic crystals), the SCNCs activation volume, ~$10^{-5}$ $b^3$, is much smaller than 1 $b^3$.

The creep deformation is however more in general regarded as the accumulation of activation events [71]. If an activation event is assumed to be a locally and kinetically controlled process, its frequency in a given volume can be expressed as $\dot{n} = \dot{n}_0 \cdot \exp(-\frac{Q_a - \sigma V}{kT})$, where $\dot{n}_0$ is the attempt frequency of activation event per unit volume, $Q_a$ is the activation energy, $\sigma$ is the applied stress on the activation volume $V$, $k$ is the Boltzmann's constant and $T$ is the absolute temperature [71]. The frequency of activation events then scales with the attempt frequency, hinting that in our case the duration of an activation event should be extremely short. The large deformations associated with superlattice dislocation movement are unlikely to occur in such short timeframes, but it is possible that a cumulation of shorter-range activation events during holding ($t$ = 1000s) ultimately leads to superlattice dislocation movement.

Indeed, C.A. Schuh found that the activation volumes corresponding to heterogeneous nucleation of dislocations (~ 0.5 $b^3$) is comparable to a vacancy volume (~0.67 $b^3$), implying that $V$ can be smaller than 1 $b^3$ and that diffusion can account for small activation volumes [71]. Compared with a NP volume in SCNCs (1.7×10$^3$ nm$^3$ here), the activation volume is still smaller by 4 orders of magnitude, indicating that NPs cannot diffuse from one site to another in the superlattice during one activation event. But considering the SCNCs' structure and constituents, activation events can also be defined as slight rearrangements of ligands, such as their extension or compaction – a phenomenon which has been found to play an important role in the plastic deformation of SCNCs [49]. Even though the $ID$ value of AP materials (~1.2 nm) suggests that ligands are interdigitated to some extent, these can still be compacted to facilitate NPs' movement. The smaller $ID$ in HT325 SCNCs, as well as the crosslinking, make the movement of NPs more difficult: the increase or decrease in inter-superlattice planes spacing in AP and HT325 materials has been found to decrease from ~20% to ~3-4% [49,63]. Yet, such slight NP movements, which are compatible with the measured $V$ values in SCNCs, can lead to a series of activation events contributing to the nucleation and propagation of dislocations, further facilitating viscoplastic deformation.

## 6. Effect of testing mode: single-loading *vs.* CSM

The main motivation behind choosing single loading-unloading mode for this creep study is to avoid the impact of the oscillations that are an integral feature of CSM testing, although CSM exhibits the advantage of less sensitivity to thermal drift [37]. Some previous studies have demonstrated that oscillations can lead to alterations in the tested materials' behavior, such as softening in metals and hardening in polymers [46], in turning impacting the measured values. While corrections for some of these effects have been proposed [43], SCNCs remain unexplored territory in this sense: The presence of an organic phase confined in nano-sized spacings can potentially make these materials particularly sensitive to oscillations-induced alterations (e.g. organic ligands rearrangement or crosslinking). The impact of oscillations in CSM on the creep behavior of SCNCs is thus here investigated, by comparing the creep displacements obtained in single loading and CSM mode, while leaving all other testing parameters unchanged.

The creep displacements associated with CSM mode are corrected for thermal drift with two methods: the one based on contact stiffness introduced in the Appendix (see Eq. (A1) – (A5)) [37], and the one based on the same procedure used here for single loading mode (own fitting of recorded thermal drift data). The comparison of all creep displacements (shown in Fig. S10) indicates that, when the same type of drift correction is applied, the creep displacements under CSM mode are smaller than those under single-loading mode. This trend becomes less pronounced in heat-treated materials. If, instead, the contact stiffness-based approach is adopted to determine creep displacements under CSM mode, a more marked difference emerges between the two testing modes.

It thus appears that in any case the CSM-associated oscillations do have an impact on the creep of SCNCs, leading to material hardening, even though the extent of this effect is not immediately quantifiable. Considering the constituents of SCNCs (organic ligands anchored to the inorganic NPs and confined in ~1 nm interfaces) and the CSM oscillations amplitude (2 nm, larger than inter-particle distances), oscillation-induced hardening can indeed occur. This is somewhat analogous to the storage modulus of polymers increasing with frequency in mechanical testing [72,73]. Further details on this analysis are given SI Section S11. The use of single-loading mode is thus confirmed to be suitable for this particular type of hybrid nanomaterials, to rule out any oscillation-induced material alterations, even though the assessment of thermal drift stays an important aspect to carefully tackle in long-term nanoindentation testing. Strain rate jump tests are a viable option for strain rate sensitivity analysis.

## Conclusions

The creep behavior of ceramic-organic supercrystalline nanocomposites (SCNCs), consisting of iron oxide NPs functionalized with oleic acid, has been investigated via nanoindentation, together with the effect of oscillations in continuous stiffness measurement (CSM) on the same creep deformation. The main findings are the following:

1) Creep occurs in SCNCs, both in presence and absence of crosslinking of the organic phase. Heat-treated (crosslinked) samples show a higher creep resistance, reflected by smaller overall creep displacements. Creep recovery also occurs, but not completely even several weeks after nanoindentation, implying that viscoelasticity and viscoplasticity are both part of the creep deformation.

2) Primary creep dominates the total creep deformation, due to a higher deformation rate at the onset of creep.

3) During quasi-secondary creep, a large stress exponent ($n > 7$) hints at the occurrence of power-law breakdown in SCNCs, an effect that is also verified by comparing threshold stress values to trigger power-law breakdown and representative under-indent stress. The increasing stress exponents with heat treatment temperature are also ascribed to the enhancement of mechanical properties induced by the crosslinking.

4) The extremely small activation volumes with respect to the superlattices Burgers vector ($\sim 10^{-5} b^3$) are interpreted as associated with slight NP movements, facilitated by rearrangement of the organic ligands in the superlattice, which are hindered when surface ligands are crosslinked.

5) Comparable strain rate sensitivity values are obtained from creep tests and strain rate jump tests.

Therefore, even though ligands only occupy a minor fraction of the SCNCs, this marked creep behavior is a clear signal of their strong influence on the nanocomposites' mechanical behavior, an effect that is tunable via crosslinking of the organic phase, which alters the inter-NP interactions. Additional control knobs to tailor the mechanical behavior of this new category of inorganic-organic nanocomposite materials can be identified in the NP size and shape and especially in the ligand density, length and characteristic backbone. Many potential research avenues can be pursued to enable the implementation of SCNCs in a plethora of different applications.

## Acknowledgements

The authors gratefully acknowledge the financial support from the Deutsche Forschungsgemeinschaft (DFG, German Research Foundation), project numbers GI 1471/1-1 and 192346071-SFB 986. The authors are thankful to Fraunhofer CAN GmbH for the TEM images of the functionalized nanoparticles, Prof. Gerold Schneider (Hamburg University of Technology) for the fruitful discussion, Dr. Jasmin Koldehoff (Hamburg University of Technology) for the assistance with nanoindentation tests, Dr. Hans Jelitto (Hamburg University of Technology) for the assistance with AFM measurements, and Dr. Emad Maawad (Institute of Materials Physics, Helmholtz-Zentrum Hereon) for assistance with the SAXS data acquisition.

## Appendix A

Due to the long holding time and low deformation rates in this creep study, the impact of thermal drift cannot be neglected. The contact stiffness is, however, less sensitive to thermal drift, and Maier et al. proposed a method to calculate indentation displacements based on contact stiffness, here summarized [37].

From Sneddon's equation,

$$S = \frac{2\beta}{\sqrt{\pi}} \cdot E_R \cdot \sqrt{A_c} \qquad (A.1)$$

where $S$ is the harmonic contact stiffness measured via nanoindentation, $\beta$ is a correction factor accounting for the indenter's shape that has a value of 1.05 for the Berkovich tip [22], $E_R$ is the reduced modulus and $A_c$ is the contact area.

The reduced modulus $E_R$ is calculated from

$$\frac{1}{E_R} = \frac{1-\nu_s^2}{E_s} + \frac{1-\nu_i^2}{E_i} \tag{A.2}$$

where the subscripts $s$ and $i$ denote sample and indenter; $E$ and $\nu$ elastic modulus and Poisson's ratio, respectively.

With Eq. (A.1) and (A.2), the contact area $A_c$ can be obtained as

$$A_c = \frac{\pi}{4\beta^2} \cdot \frac{S^2}{E_R^2} \tag{A.3}$$

The contact area function of indenter is calibrated on silica, and it is found to be:

$$A_c = 24.5h_c^2 + 249.52h_c + 173.4h_c^{0.5} + 78.91h_c^{0.25} + 51.87h_c^{0.125} \tag{A.4}$$

By numerically solving eq. (A.4), the value of contact depth $h_c$ can be obtained. The total displacement can then be calculated as

$$h_{csm} = h_c + \varepsilon \cdot \frac{P}{S} \tag{A.5}$$

where $\varepsilon$ is a constant depending on the geometry of the indenter, which has a value of 0.75 for Berkovich tip [74].

The difference between $h_{csm}$ and $h_{meas}$ (raw nanoindentation data) is ascribed to the effect of thermal drift:

$$h_{thermal} = h_{meas} - h_{csm} \tag{A.6}$$

The time derivative of $h_{thermal}$ is the thermal drift:

$$\lambda = dh_{thermal}/dt \tag{A.7}$$

We compare the resulting thermal drift with the one resulting from alternative approaches. These are: (i) the thermal drift data measured by the nanoindenter itself, as fitted by the device or with an own, more accurate, fit, and (ii) the thermal drift data resulting from indenting reference materials for which the drift is either known or negligible, namely silica and sapphire. All these options are evaluated for their suitability to our tests, and the most appropriate is then selected and applied to the creep data.

More specifically, the following thermal drift assessment methods are implemented and evaluated. The thermal drift rates obtained with the different methods are compared in Fig. A1.

Under CSM mode:
a) drift calculated based on the measured contact stiffness (calc_csm in the plot);
b) drift provided by the nanoindenter based on the dedicated segment in the tests (measured_csm);
c) own fitting (fitting_csm) of the data of (b);
d) data extracted by silica tests;
e) data extracted by sapphire tests.

Under single loading mode:
f) data provided by the nanoindenter based on the dedicated segment in the tests (measured_single loading);
g) own fitting (fitting_single loading) of the data of (f).

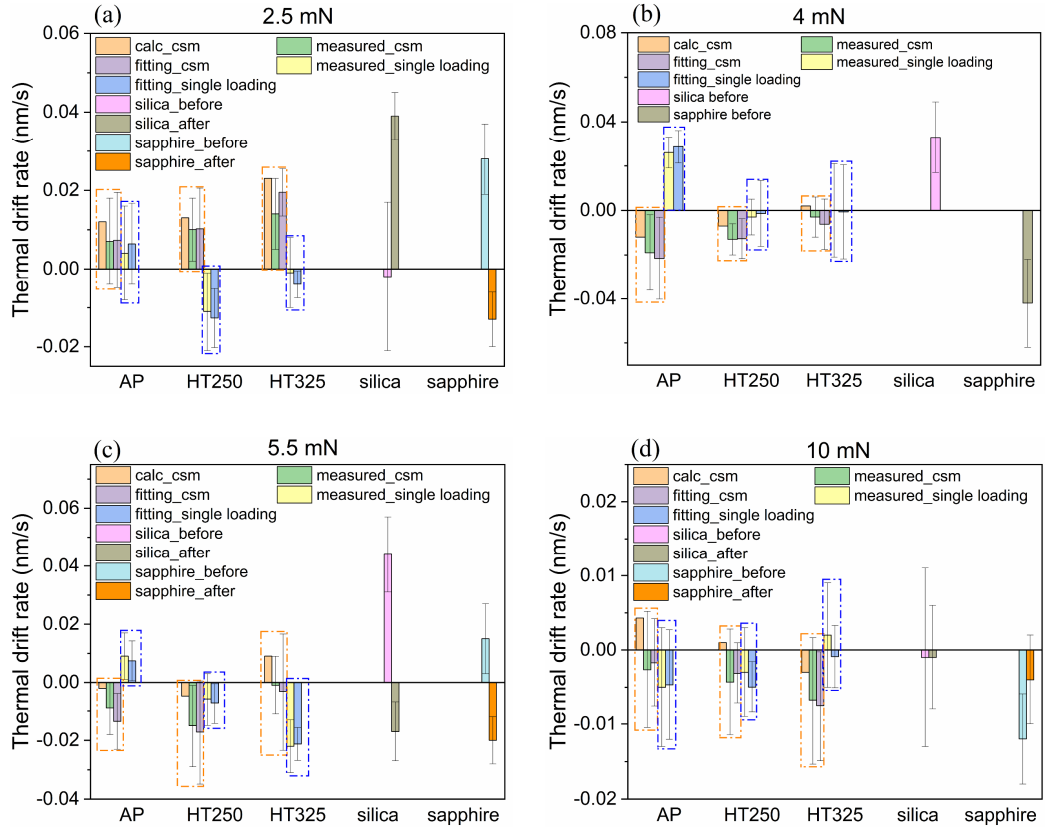

**Fig. A1.** Comparison of thermal drift rates obtained from different methods: (a) 2.5 mN; (b) 4 mN; (c) 5.5 mN; (d) 10 mN. The thermal drift rates measured under CSM mode are marked by orange dashed rectangles, while those under single loading mode are marked by blue dashed rectangles. The other data is relative to silica and sapphire. Note that the latter are missing for the after-creep stage in the 4 mN tests, due to an unforeseen interruption of the testing cycles after the creep measurements.

Under CSM mode, the drift values relative to silica and sapphire are not consistent before (silica/sapphire_before in the plot) and after creep tests (silica/sapphire_after in the plot) for the 2.5 mN and 5.5 mN tests, and they are not comparable to those from Eq. (A.7). This drift data is thus discarded. The drift values based on the nanoindenter measurement and on own linear fitting are comparable, and they are also comparable to those from Eq. (A.7), indicating that this data can be employed for the correction.

Under single loading mode, the drift values from measurement of nanoindentation and own linear fitting are also comparable, but the drift values from Eq. (A.7) under CSM mode significantly deviate from these in certain tests. Therefore, the value from Eq. (A.7) cannot be employed here to correct the raw data obtained under single loading mode.

Based on these considerations, the data obtained from the thermal drift measurements will be adopted for the correction of the raw data measured in single loading mode, with an own fitting applied to improve accuracy, while for CSM mode both the contact-stiffness based method (a) and the same procedure adopted to correct single loading data (c) are applied.

# Supplementary Information

Cong Yan[1], Büsra Bor[2], Alexander Plunkett[2], Berta Domènech[2], Verena Maier-Kiener[3], Diletta Giuntini[1,2,*]

[1] Department of Mechanical Engineering, Eindhoven University of Technology, The Netherlands

[2] Institute of Advanced Ceramics, Hamburg University of Technology, Germany

[3] Department of Materials Science, Montanuniversität Leoben, Austria

## 1. TEM and SAXS analyses

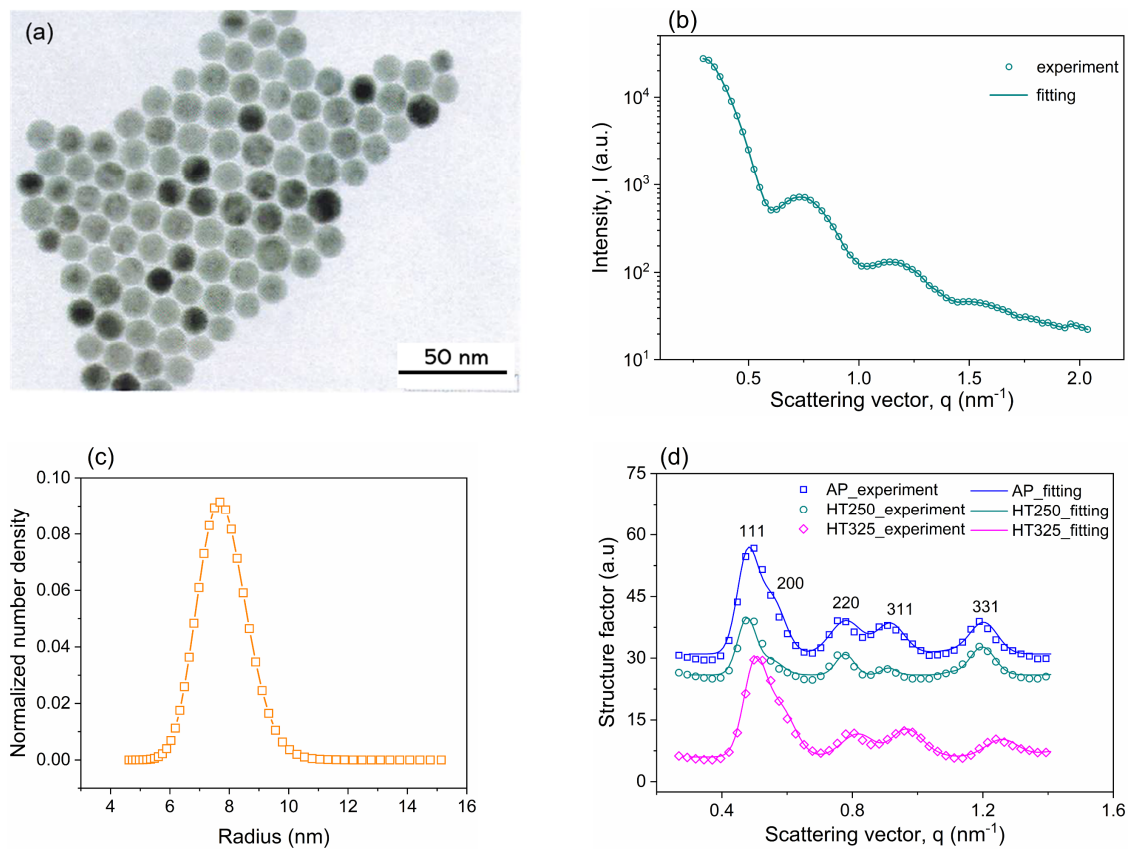

**Fig. S1.** The characterization of nanoparticles (NPs) and superlattice via TEM and SAXS. (a) TEM image of OA-functionalized NPs before self-assembly (provided by the supplier, Fraunhofer CAN GmbH, Hamburg, Germany); (b) SAXS scattering curve of the initial suspension; (c) NP size distribution obtained from SAXS, analyzed by number density-based fitting and lognormal adjustment; (d) Structure factor of supercrystalline nanocomposites (SCNCs) obtained from SAXS, for AP, HT250 and HT325 samples. The structure factor is obtained as the intensity divided by the form factor. A baseline correction of the structure factor is performed, to determine the peak position more accurately. The curves of AP and HT250 are shifted vertically for clarity.

[*] Corresponding author: d.giuntini@tue.nl

The TEM image in Fig. S1(a) reveals an almost uniform NP size distribution, with the radius of 7.5 ± 0.7 nm. This is in good agreement with the NP size obtained via small-angle X-ray scattering (SAXS), analyzed as detailed elsewhere [1]. In Fig. S1(c), the NP radius is determined as 7.4 ± 0.8 nm from the scattering curve of suspension shown in Fig. S1(b), which is assumed to be the radius of iron oxide core, without the oleic acid (OA) functionalization, since the latter is significantly less detectable by X-rays, and this data is collected when the NPs are still in suspension [2]. Face-centered-cubic (FCC) superlattice is revealed for all SCNC samples, based on the structure factor displayed in Fig. S1(d), with superlattice constants of 22.8 ± 0.04 nm, 22.9 ± 0.03 nm and 21.9 ± 0.12 nm for AP, HT250 and HT325, respectively. The slightly higher superlattice constant of HT250 compared to that of AP, is within the SAXS measurement error (0.2 nm).

## 2. Organic content (TGA analysis)

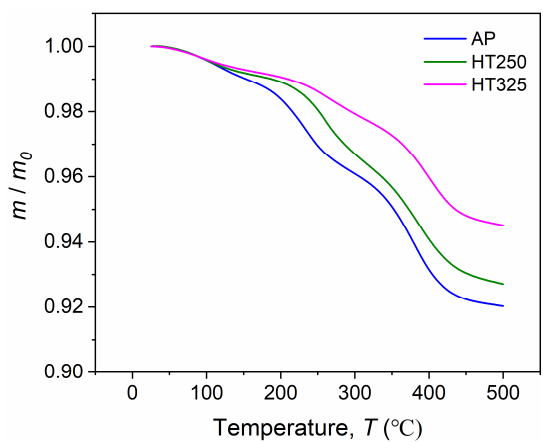

**Fig. S2.** Mass variation of different materials as measured by TGA. The ratio: $m/m_0$ indicates the mass loss with increasing temperature, where $m_0$ is the initial mass and $m$ is the real-time mass during the measurement.

The mass loss of three different materials upon heating until 500 °C is shown in Fig. S2 (the loss under higher temperatures is not considered as relevant for the organic content assessment, because oxygen from the magnetite leads to the blowout of CO and $CO_2$, and thus an overestimation of organic content [3]). The AP sample exhibits the most significant mass loss, followed by HT250 and HT325. The TGA curves do not decrease monotonically, but stepwise, which is related to different mechanisms regarding the decomposition of ligands with increasing temperature [1].

## 3. Estimation of NP monolayer ligand grafting density

The grafting density is defined as the number of ligand molecules per NP unit surface [4], with the NP assumed to be spherical, i.e.

$$\rho_{OA} = \frac{N_{OA}}{S_{NP}} = \frac{N_{OA}}{\pi d^2 \cdot N_{NP}} \quad (1)$$

where $\rho_{OA}$ is the grafting density, $N_{OA}$ is the number of ligands (oleic acid) on the NPs surface, $S_{NP}$ is the surface area of all NPs, $N_{NP}$ is the number of NPs and $d$ is the NP diameter.

The number of oleic acid on the surface of NPs can be estimated from

$$N_{OA} = N_A \cdot \frac{m_{OA}}{M_{OA}} = N_A \cdot \frac{w_{OA} \cdot m}{M_{OA}} \quad (2)$$

where $N_A$ is Avogadro number, $m_{OA}$ is the mass of oleic acid, $M_{OA}$ is the molar mass of oleic acid, $m$ is the sample's mass and $w_{OA}$ is the weight fraction of oleic acid.

The number of NPs ($N_{NP}$) can be obtained from:

$$N_{NP} = \frac{N_{Fe}}{N_{NP,Fe}} \quad (3)$$

where $N_{Fe}$ is the number of Fe atoms in the sample while $N_{NP,Fe}$ is the number of Fe atoms in one NP.

The total number of Fe atoms in the sample is calculated as follows:

$$N_{Fe} = \frac{m \cdot w_{Fe_3O_4}}{M_{Fe_3O_4}} \cdot N_A \cdot 3 = 3 N_A \cdot \frac{m \cdot (1 - w_{OA})}{M_{Fe_3O_4}} \quad (4)$$

where $w_{Fe_3O_4}$ is the weight fraction of Fe$_3$O$_4$ and $M_{Fe_3O_4}$ is its molar mass.

The volume of magnetite unit cell is given as $V_{UC}$=0.5915 nm³ with 8 formula units [1], thus 24 Fe atoms in each unit cell. Therefore, the number of Fe atoms in one NP is:

$$N_{NP,Fe} = 24 \cdot \frac{V_{NP}}{V_{UC}} = \frac{4\pi d^3}{V_{UC}} \quad (5)$$

Given Eq. (1) – (5), the grafting density is obtained as follows:

$$\rho_{OA} = \frac{w_{OA}}{M_{OA}} \cdot \frac{M_{Fe_3O_4}}{1 - w_{OA}} \cdot \frac{4d}{3V_{UC}} \quad (6)$$

With $w_{OA}$ = 8.0%, $M_{OA}$ = 282 g/mol, $M_{Fe_3O_4}$ = 232 g/mol, $d$ = 14.9 nm (from SAXS), the experimental grafting density for monolayer ligand is determined as $\rho_{OA}$ = 2.4 molecules/nm².

## 4. Measurement of elastic modulus and hardness

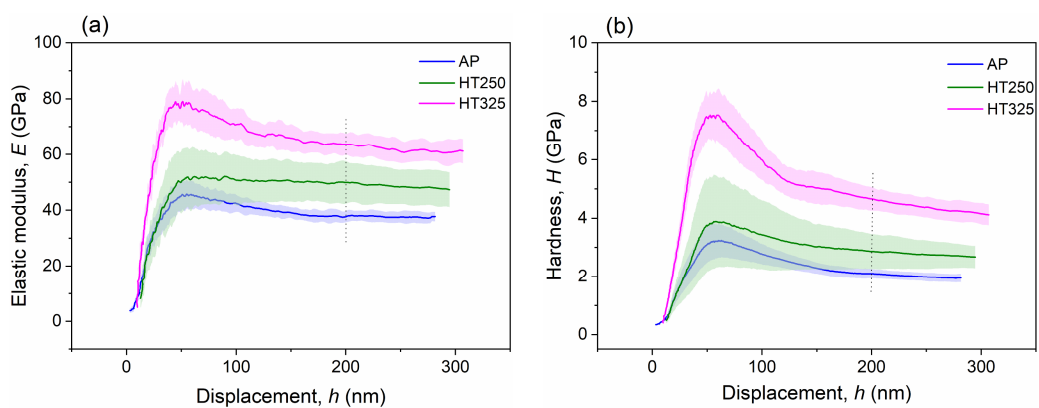

**Fig. S3.** Nanoindentation-measured mechanical properties *vs* displacement of SCNCs: a) elastic modulus and b) hardness. Only the data beyond 200 nm (indicated by the dashed line) is averaged to determine elastic modulus and hardness. The shaded bands indicate the standard deviation.

## 5. SEM images of 10 mN indents

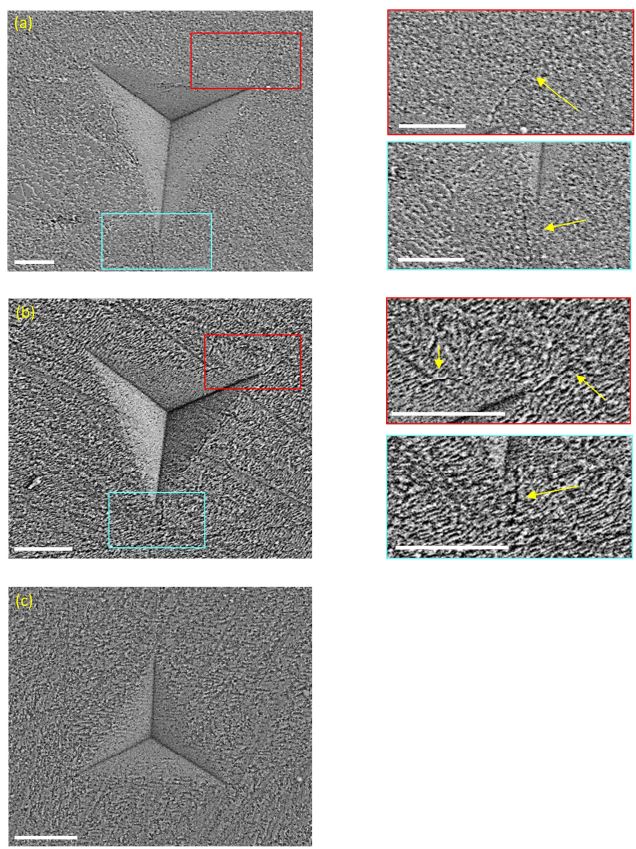

**Fig. S4.** SEM micrographs of indents obtained at 10 mN in: (a) AP sample; (b) HT250 sample; (c) HT325 sample. The indentation-induced cracks are highlighted, as indicated in the magnified view on the right. Scale bars: 1μm.

For AP and HT250 samples, only some minor cracks can be observed around indents, while chipping is not detected. No cracks nor chipping are found in the HT325 samples, due to the higher fracture toughness [5].

## 6. Evolution of creep and backcreep displacements

The evolution of creep displacements under 2.5, 4 and 10 mN is displayed in Fig. S5(a), (c), (e), respectively. The creep displacements first increase sharply and then slow down. The sample with higher heat treatment temperature (325 ºC) shows smaller creep displacements, ascribed to higher creep resistance resulting from crosslinking. The evolution of backcreep displacements in 2.5, 4 and 10 mN tests is displayed in Fig. S5(b), (d), (f), respectively. They drop quickly at the onset of holding, and subsequently slow down.

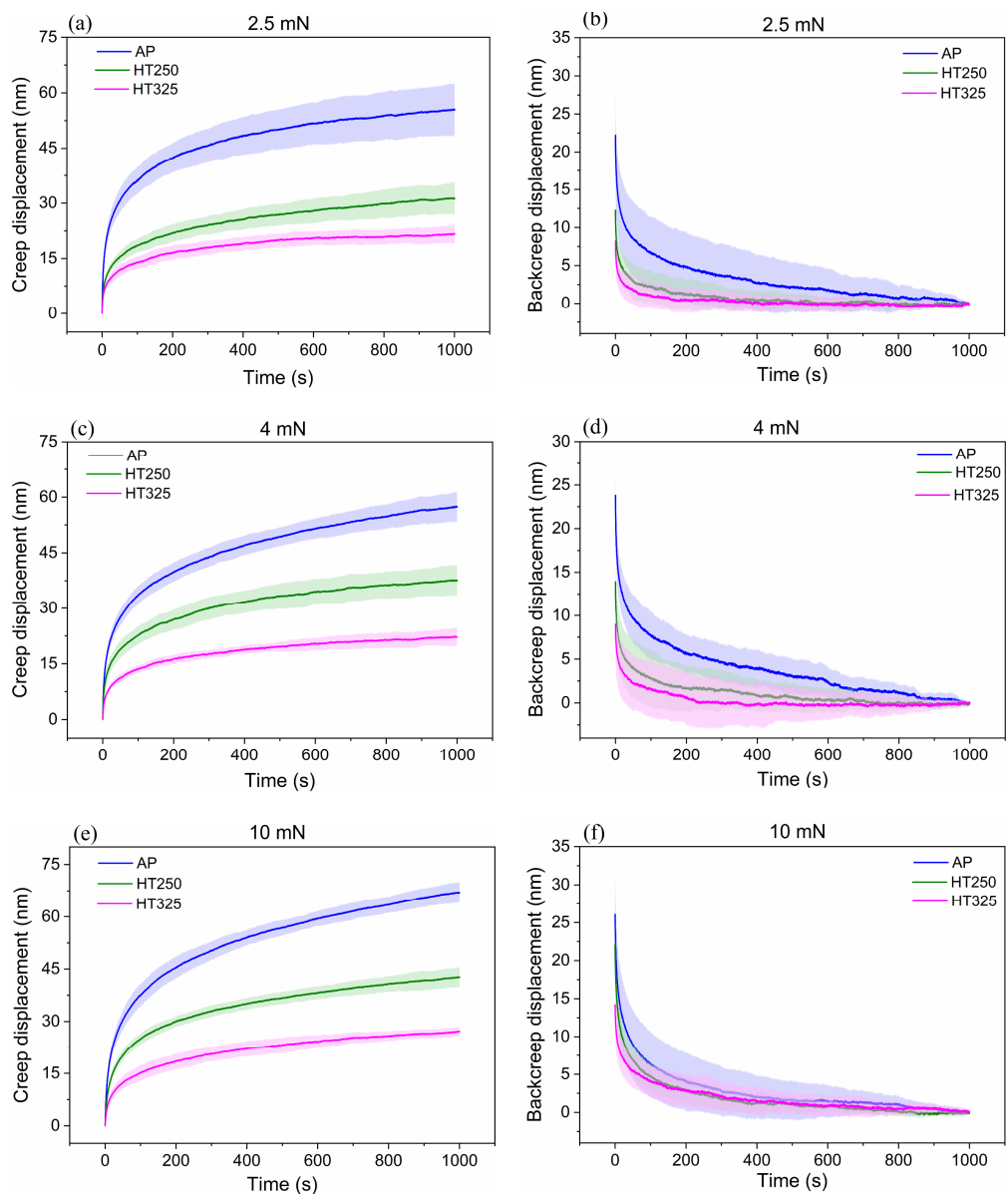

**Fig. S5.** Evolution of creep and backcreep displacements under different loads: (a) creep under 2.5 mN; (b) backcreep in 2.5 mN tests; (c) creep under 4 mN; (d) backcreep in 4 mN tests; (e) creep under 10 mN; (f) backcreep in 10 mN tests; The shaded bands indicate the standard deviation.

## 7. Stress evolution and applicability of KWW model

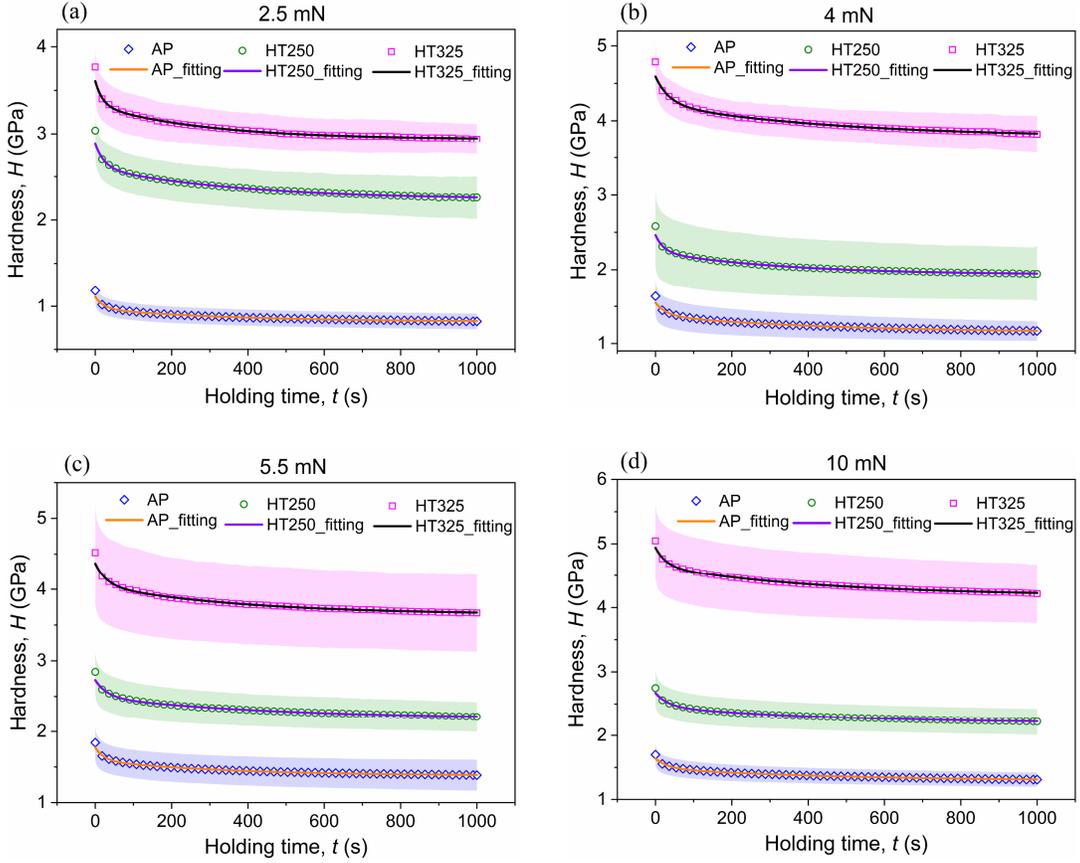

**Fig. S6.** Evolution of hardness (representative of stress in the Kohlrausch-Williams-Watts model) during holding: (a) 2.5 mN; (b) 4 mN; (c) 5.5 mN; (d) 10 mN; The standard deviation is indicated by shaded bands. A function with double exponential terms is employed to fit the experimental data, yielding a very high correlation coefficient ($R^2 > 0.99$). The scatters are from experimental data while solid lines indicate the fitting.

The viscoelastic stress evolution during creep can be captured by the Kohlrausch-Williams-Watts (KWW) model [6–8], as follows:

$$\sigma = \sigma_0 \exp\left[-(t/\tau)^\beta\right] \tag{7}$$

where $\sigma_0$ is an instantaneous stress, $\tau$ is the time constant and $\beta$ is a power law exponent with its value in the range $0 < \beta < 1$. Here, the stress is approximated by hardness (mean pressure) obtained from Eq. (3) in the main text, as displayed in Fig. S6. However, the KWW model fails to match the stress evolution in time of SCNCs, while a function with double exponential terms ($y = y_0 + a_1 * \exp\left(-\frac{x}{t_1}\right) + a_2 * \exp\left(-\frac{x}{t_2}\right)$) fits the results very well ($R^2 > 0.99$), as shown by the solid lines. The same phenomenon was also found in enamel, where viscoelasticity and viscoplasticity both exist [9].

## 8. Assessment of deformation during primary creep

**Table S1.** The ratio of strain during primary creep to the total one under the four different loads.

|       | 2.5 mN | 4 mN  | 5.5 mN | 10 mN |
|-------|--------|-------|--------|-------|
| AP    | 88.7%  | 85.9% | 86.3%  | 85.4% |
| HT250 | 85.4%  | 87.0% | 85.9%  | 86.7% |
| HT325 | 88.8%  | 87.1% | 84.0%  | 85.4% |

The ratio is defined as the strain during primary creep divided by the total strain, as obtained from Fig. 4 in the main text. All the ratios are larger than 80%, indicating that deformation during primary creep occupies a large amount of creep deformation. No clear trend based on the material type is observed.

## 9. Correlation between creep tests and strain-rate jump tests

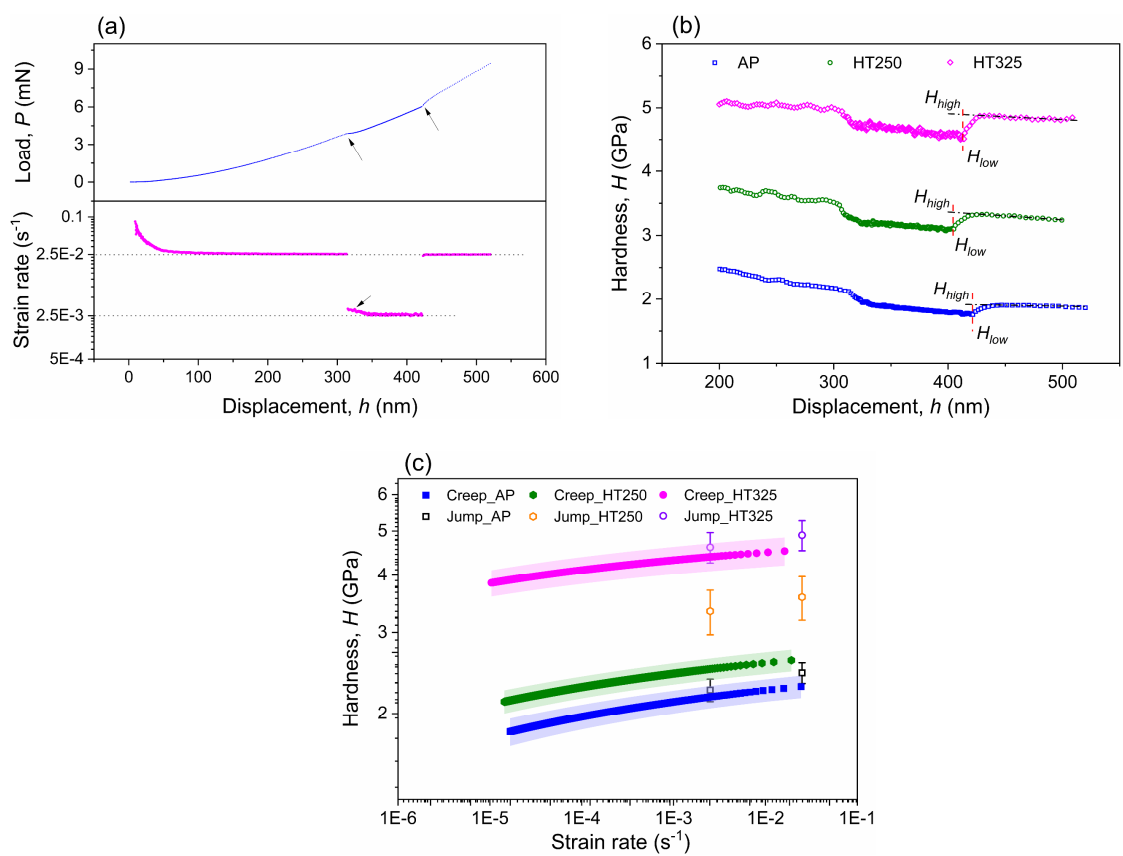

**Fig. S7.** Strain rate jump tests. (a) Load-displacement curve of strain rate jump test on AP SCNC (only the loading segment is shown). The lower plot illustrates the corresponding variation of strain rate. Two deflections (indicated by arrows) appear on the load-displacement curve due to the jump of strain rate. (b) The evolution of hardness, with displacement, in the 200 - 500 nm range, highlighting the variation of $H$ caused by the strain rate jump. $H_{low}$ and $H_{high}$ indicate the hardness under low and high strain rate, respectively. (c) Relationship between hardness and strain rate in a log-log plot, as resulting from creep (solid symbols) and strain-rate jump (hollow symbols) tests.

The hardness and strain rate associated with creep and strain rate jump tests are plotted in the same figure for comparison. The data from the strain rate jump test falls in the proximity of creep test master curves for AP and HT325 SCNCs, while the same agreement is not found for HT250 materials. This discrepancy is to be investigated further, but it still reasonable to conclude that the two methods lead to results that are not significantly different.

## 10. Activation volume

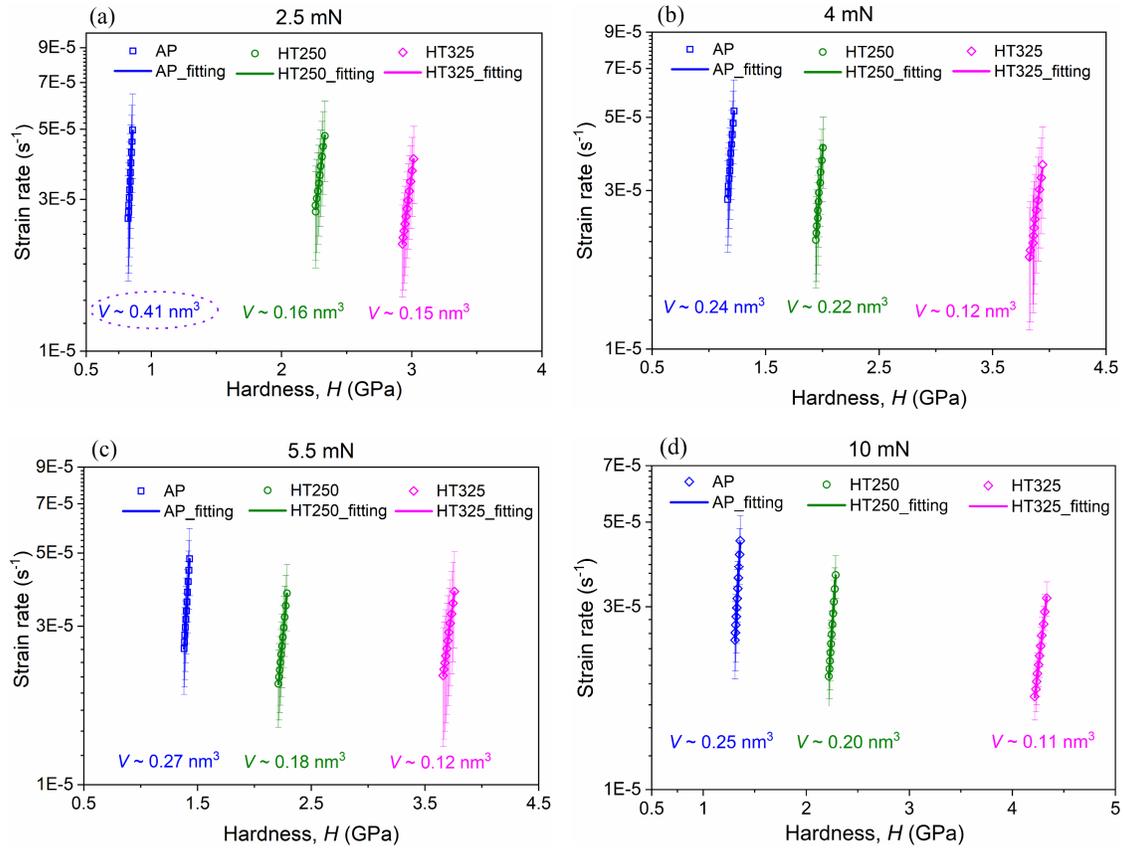

**Fig. S8.** Relations of logarithmic strain rate vs. linear hardness during creep holding, from which activation volumes are determined: (a) 2.5 mN; (b) 4 mN; (c) 5.5 mN; (d) 10 mN. For the AP sample, an outlier is found in the activation volume in the 2.5 mN tests, which is then excluded in the average estimations.

Based on the Eq. (5) in the main text, the activation volume ($V$) is identified as the slope of logarithmic strain rate vs. linear hardness plot, which is also marked in the respective plots. The activation volume is found to exhibit a decreasing trend with rising heat-treatment temperature.

## 11. Influence of oscillations in continuous stiffness measurement (CSM)

The superimposition of oscillations in CSM mode enables the continuous measurement of contact stiffness, allowing to determine the displacement via contact stiffness-based approach, as mentioned in the main text. To ensure that the contact stiffness values obtained in CSM mode are reliable, a check on the ratio of elastic modulus to hardness ($E/H$) and on the phase angle is performed. A low $E/H$ ratio implies negligible plastic deformation during CSM cycles, which can be a cause of contact stiffness underestimation [10]. In [10], it is reported that contact stiffness can be reliably measured if $E/H$ < 40. For the SCNCs studied here, the $E/H$ ratio is even lower (~19 for AP materials, ~18 for HT250 and ~14 for HT325), suggesting that the contact stiffness bias is not affecting the measurements. The phase angle, defined as the angle between load and displacement during sinusoidal deformation imposed by the harmonic load [11], is also monitored, because a stable phase angle is an indicator of the reliability of the contact stiffness measurements, and in general a shift below ~5 degrees is considered acceptable [10]. The phase angle during holding is monitored for all four different loads, and in all cases an almost identical trend is observed, therefore only results relative to the 4 mN case are shown in Fig. S9. The phase angle remains stable during the entire holding, within a shift of only 2 degrees, which guarantees the suitability of the contact stiffness-based approach to assess displacements in CSM mode.

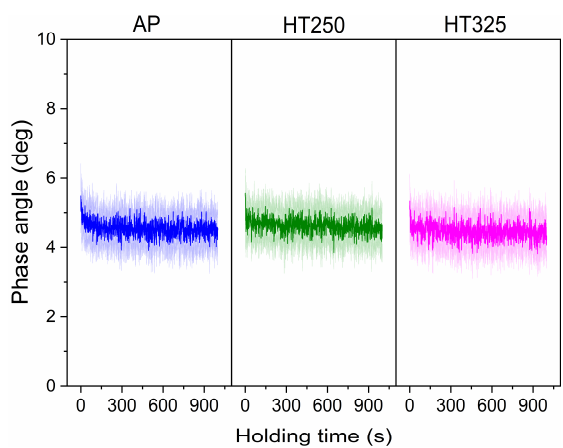

**Fig. S9.** Variation of phase angle during holding under 4 mN. The standard deviation is indicated by the bands.

Additionally, the thermal drift correction used to correct raw data in single loading mode is also applied to CSM data. Creep displacements between single loading and CSM mode is compared in Fig. S10 to explore the potential influence of oscillations on the mechanical behavior of SCNCs. Smaller creep displacements are found in CSM mode, no matter which way is used to determine the displacement without thermal drift effects. This trend becomes less pronounced for heat-treated samples. It is worth mentioning explicitly here that an out-of-trend creep displacement (43.3 nm) had been observed in one region of the AP sample during 4 mN CSM tests, and that these measurements were thus repeated in three additional different regions, leading instead to the reproducible creep displacements (~ 30 nm) that are displayed in the plot. Organic-rich clusters can occur within SCNCs, leading to locally soft responses. The smaller creep displacements in CSM mode hint at hardening phenomena.

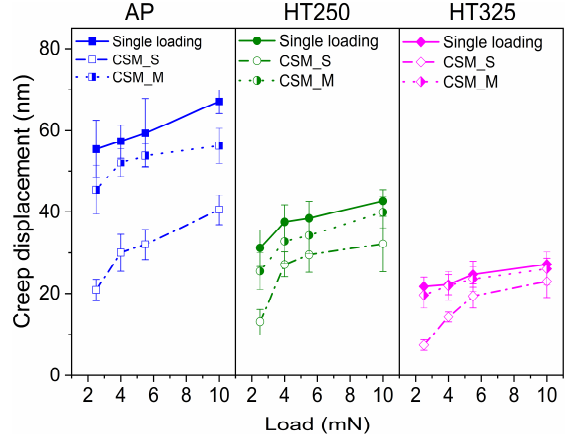

**Fig. S10.** Creep displacements under single loading and CSM mode. In addition to creep displacements from contact stiffness-based approach (CSM_S), those directly obtained from correction based on thermal drift measurement are also shown in the plot (CSM_M) for CSM mode. The latter drift correction approach is also the one used for single-loading data.

Two main aspects are distinctive of the CSM mode compared to the single loading-unloading tests: the magnitude of the superimposed harmonic load, leading to an overall slightly larger maximum applied load, and its dynamic character (periodic oscillations). The amplitudes are given in detail in Table S2, showing that the harmonic loads applied by the CSM method increase with both nominally applied load and heat-treatment temperature. Since the SCNCs tend to be compacted in nanoindentation (reduction of the inter-NP distance by increase organic interdigitation and potential filling up of the interstitial sites) [5,12], increasing applied loads can imply larger compaction and thus higher harmonic loads needed to achieve the same oscillation amplitude. The increasing harmonic loads when shifting from AP to HT250 and HT 325 SCNCs are also correlated with the crosslinking-induced material stiffening (see Table 1 in the main text). The ratios of harmonic load to maximum applied load, also summarized in Table S2, are in the 2 – 5% range, a minor contribution that cannot account for the difference in creep displacements in Fig. S10.

**Table S2.** Harmonic loads under CSM mode and corresponding ratios with respect to maximum applied load.

|       | Absolute value (µN) | | | | Ratio | | | |
|-------|--------|-------|--------|-------|--------|-------|--------|-------|
|       | 2.5 mN | 4 mN  | 5.5 mN | 10 mN | 2.5 mN | 4 mN  | 5.5 mN | 10 mN |
| AP    | 111.4  | 149.9 | 160.5  | 199.9 | 4.5%   | 3.8%  | 2.9%   | 2.0%  |
| HT250 | 120.0  | 157.3 | 166.7  | 204.4 | 4.8%   | 3.9%  | 3.0%   | 2.0%  |
| HT325 | 130.8  | 167.0 | 175.8  | 242.1 | 5.2%   | 4.2%  | 3.2%   | 2.4%  |

As for the influence of the dynamic character of the load, this can be assessed into the framework of strain rate sensitivity. It has indeed been reported that the difference in strain rates between the quasi-static single loading mode (0.005 s$^{-1}$) and the dynamic CSM mode (0.05 s$^{-1}$) can lead to different values of hardness for strain rate sensitive materials [13]. The harmonic displacement (2 nm) is indeed imposed in a very short timestep (half loading cycle, corresponding to 1/180 s in this study), resulting in an extremely high strain rate, typically 5 orders of magnitude higher than that under single loading mode in this study. SCNCs are strain rate sensitive materials (*m*:

0.04 ~ 0.06), as Fig. 5(e) has shown. One can adopt $m$ to estimate the stress increase associated with the oscillations, starting from the definition of strain rate sensitivity itself. From Eq. (4) in main text, we can rewrite it to obtain the stress under different strain rates as follows, assuming unchanging $m$,

$$\frac{\sigma_1}{\sigma_2} = \left(\frac{\dot{\varepsilon}_1}{\dot{\varepsilon}_2}\right)^m \tag{8}$$

where subscript "1" denotes under CSM mode while "2" denotes under single loading mode. This estimation is applied to HT325 materials tested at 2.5 mN. The harmonic amplitude is approximated to follow a linear change with time in order to simplify the calculation of $\dot{\varepsilon}_1$, which has a value of 1.8 s$^{-1}$, while the strain rate $\dot{\varepsilon}_2$ is estimated as 1.6×10$^{-5}$ s$^{-1}$ based on the end of the corresponding holding segment. With these two strain rates and the strain rate sensitivity as given in Fig. 5(e), the stress ratio ($\sigma_1/\sigma_2$) is determined as 1.74, suggesting that the stress required for material flow increases by 74% during the loading cycles of oscillations. An even larger stress increase is found in 2 other samples, i.e. 102% for AP and 77% for HT250, since they feature a higher $m$. It is thus reasonable to attribute the observed hardening to this effect.